\documentclass[twocolumn,groupedaddress,superscriptaddress]{revtex4}
\usepackage{lmodern}

\usepackage[T1]{fontenc}
\usepackage[utf8]{inputenc}
\usepackage{color}
\usepackage{amsmath}
\usepackage{amssymb}
\usepackage{graphicx}
\usepackage[unicode=true,pdfusetitle,
 bookmarks=true,bookmarksnumbered=false,bookmarksopen=false,
 breaklinks=false,pdfborder={0 0 1},backref=section,colorlinks=true]
 {hyperref}
\hypersetup{
 linkcolor=black,hyperindex=true}
\usepackage{breakurl}

\makeatletter
\@ifundefined{textcolor}{}
{%
 \definecolor{BLACK}{gray}{0}
 \definecolor{WHITE}{gray}{1}
 \definecolor{RED}{rgb}{1,0,0}
 \definecolor{GREEN}{rgb}{0,1,0}
 \definecolor{BLUE}{rgb}{0,0,1}
 \definecolor{CYAN}{cmyk}{1,0,0,0}
 \definecolor{MAGENTA}{cmyk}{0,1,0,0}
 \definecolor{YELLOW}{cmyk}{0,0,1,0}
}

\usepackage{amsfonts}\usepackage{xcolor}\usepackage{setspace}%\usepackage{cite}
\usepackage{indentfirst}\usepackage{epstopdf}

\newcommand{\ket}[1]{|#1\rangle} % ket-
\newcommand{\identity}{{\bf 1}_{2 \times 2}} % \newcommand{\identity}{{\bf 1}\!{\sf l}}

\makeatother

\begin{document}

\title{Splitting of the topologically-protected Dirac cone without breaking time reversal symmetry}

%\title{Splitting of the topologically-protected Dirac cone at the interface between topological insulator and semiconductor}

\author{L. Seixas}

%\email{lseixas@if.usp.br}

\affiliation{Instituto de F\'{i}sica, Universidade de São Paulo, CP 66318, 05315-970,
São Paulo, SP, Brazil}

\affiliation{Department of Physics, Applied Physics, and Astronomy, Rensselaer
Polytechnic Institute, Troy, New York 12180, USA}

\author{D. West}

\email{damienwest@gmail.com}

\affiliation{Department of Physics, Applied Physics, and Astronomy, Rensselaer
Polytechnic Institute, Troy, New York 12180, USA}

\author{A. Fazzio}

%\email{fazzio@if.usp.br}

\affiliation{Instituto de F\'{i}sica, Universidade de São Paulo, CP 66318, 05315-970,
São Paulo, SP, Brazil}

\author{S. B. Zhang}

%\email{zhangs9@rpi.edu}

\affiliation{Department of Physics, Applied Physics, and Astronomy, Rensselaer
Polytechnic Institute, Troy, New York 12180, USA}

\date{\today}

\begin{abstract}
Topological insulators (TIs) are a new class of matter characterized
by the unique electronic properties of an insulating bulk and metallic
boundaries arising from non-trivial bulk band topology. While the
surfaces of TIs have been well studied, the interface between TIs
and semiconductors may not only be more technologically relevant but
the interaction with non-topological states may fundamentally alter
the physics. Here, we present a general model to show that such an
interaction can lead to spin-momentum locked non-topological states,
the Dirac cone can split in two, and the particle-hole symmetry can
be fundamentally broken, along with their possible ramifications.
Unlike magnetic doping or alloying, these phenomena occur without
topological transitions or the breaking of time reversal symmetry.
The model results are corroborated by first-principles calculations
of the technologically relevant Bi$_{2}$Se$_{3}$ film van der Waals
bound to a Se-treated GaAs substrate. 
\end{abstract}
\maketitle

Topological Insulators (TIs) are somewhat unique in the history of
condensed matter physics as both the nature of these materials and
many of their associated exotic properties were predicted on purely
theoretical grounds before experimental evidence of their existence
was found\cite{PhysRevLett.95.226801,PhysRevLett.95.146802,Bernevig15122006}.
The topology of an insulator is associated with the continuous deformation
of its Hamiltonian; if it can be continuously deformed to yield the
band structure of another insulating system, while maintaining a bandgap
throughout the deformation, then they belong to the same topological
group and have the same associated $\mathbf{Z}_{2}$ topological invariant\cite{PhysRevLett.95.146802}.
The relatively small bandgap and large spin-orbit interaction (SOI)
in TIs leads to the nontrivial topology and a $\mathbf{Z}_{2}$ invariant
which is distinct from normal insulators (NIs) such as GaAs, Si, or
even vacuum. Although the bulk properties of TIs are similar to ordinary
insulators, the interface between two materials with different topological
invariants leads to the emergence of a topologically-protected metallic
state which is localized to the interface and has a linear Dirac-cone-like
dispersion relation, as described in the Kane—Mele model\cite{PhysRevLett.95.226801,hasan2010colloquium,qi2011topological}.

Although most current studies are focused on TI surfaces \cite{Chen10072009,hsieh2009tunable,alpichshev2010stm,PhysRevLett.110.026602,hsieh2009observation,li2014electrical},
this represents only a special case of all possible interfaces, i.e.,
the interface with vacuum. Despite being less studied, real interfacial
states are not only unavoidable at the substrate, but may be even
more advantageous for applications than utilization of the surface
states. Interfacial states are protected from the environment, indispensable
for creating edge-state electrical circuits, and the formation of
periodic heterostructures of TIs and NIs may provide numerous channels
for topological current, potentially obtaining bulk-like current densities
while maintaining topological protection. Similar to the surface,
it is believed that interfacial topological states are spin-momentum
locked, where the spin polarization of the electron is locked in plane
and perpendicular to the crystal momentum, $k$, leading to a number
of potential applications \cite{li2014electrical,fan2014magnetization}.
They may find great use in spintronics, as the spin current can now
be controlled without magnetism but instead through the application
of $E$-fields leading to new architectures for spin based transistors\cite{awschalom2009spintronics,pesin2012spintronics,zutic2004spintronics}.
Furthermore, the weak antilocalization of the Dirac-state is robust
against disorder and localized non-magnetic perturbations cannot lead
to backscattering of the electron, yielding dissipationless transport.

The current lack of attention on interface states is not due to their
lack of importance, but instead on the experimental difficulty in
their characterization. As they are beyond the scope of standard surface
sensitive techniques, little is experimentally known and they have
only recently been detected\cite{yoshimi2014dirac}. While the interfacial
Dirac cone is generally believed to have essentially identical characteristics
to that of the surface, we note that due to the large difference in
ionization energies among promising TIs (such as the pnictogen chalcogenides)
and ordinary semiconductors, other localized states at the interface
may greatly alter the interface properties.

\begin{figure}
\centering \includegraphics[width=0.45\textwidth]{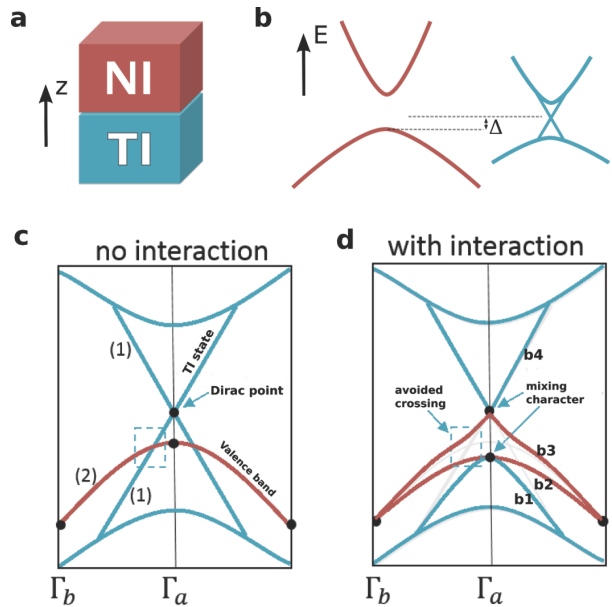} \caption{\textbf{TI/NI interface effective Hamiltonian model.} \textbf{(a)}
Schematic representation of the type-II interface between a topological
insulator (TI) and a normal insulator (NI). \textbf{(b)} Band structures
of semiconductor (red) and topological insulator (blue). The $\Delta$
parameter in the Hamiltonian model is the VBM energy relative to the
Dirac point. \textbf{(c)} Depiction of the topological interface state
and associated Dirac point and the bulk semiconductor derived bands
within the topological band gap in the absence of mutual interaction.
The degeneracy of the bands are labeled and the black dots indicate
points which are required to be doubly degenerate due to the preservation
of TRS. \textbf{(d)} shows the effect to (c) when interaction is turned
on. The anticrossing between the doubly degenerate semiconductor band
and the non-degenerate TI state requires band change character from
semiconductor to TI. Neither of the double degeneracies at $\Gamma_{a}$
can cleanly be labeled as the Dirac point as both consist of the crossings
of state which change from TI to semiconductor character.}

\label{fig:Fig1} 
\end{figure}

Herein, we construct a simple model to describe the interaction of
the topological interface state with non-topological states at the
interface. Although such situations may arise due to defects in the
near interface region, they are a general consequence of misalignment
of the band edges of the TI and the NI. Such misalignment leads to
a large transverse $E$-field in the vicinity of the interface which
can localize bulk semiconductor bands to the near interfacial region
-- resulting in substantial interaction with the topological state.
In this model, the pristine topological insulator states at the interface
are described by a massless Dirac cone with helical (spin-momentum-locked)
spin textures represented by the effective surface Hamiltonian, $v_{F}\hslash(\mathbf{k}\times\boldsymbol{\sigma})\cdot\hat{{\bf z}}$.
Given that in general TIs have a narrow bulk bandgap, the simplest
model would be to consider that the Dirac cone interacts with only
a single band of the semiconductor, as shown in Fig. \ref{fig:Fig1}(b-c),
which is described by a spin-degenerated parabola, $\left(\frac{\hslash^{2}}{2m^{*}}|\mathbf{k}|^{2}+\Delta\right)\identity$.
Using the basis $\ket{\psi}=\left(\ket{\psi_{{\rm TI}}^{\uparrow}}\ \ \ket{\psi_{{\rm TI}}^{\downarrow}}\ \ \ket{\psi_{{\rm NI}}^{\uparrow}}\ \ \ket{\psi_{{\rm NI}}^{\downarrow}}\right)^{T}$,
where $\ket{\psi_{{\rm TI}}^{\uparrow,\downarrow}}$ are the topological
insulators spinors and $\ket{\psi_{{\rm NI}}^{\uparrow,\downarrow}}$
are the semiconductor spinors, we can write the interaction between
the topological and semiconductor states in $k$-space as follows,
\begin{equation}
H({\bf k})=\begin{pmatrix}v_{F}\hslash({\bf k}\times\boldsymbol{\sigma})\cdot\hat{{\bf z}} & V_{{\rm int}}\\
V_{{\rm int}}^{\dagger} & \left(\frac{\hslash^{2}}{2m^{*}}|\mathbf{k}|^{2}+\Delta\right)\identity
\end{pmatrix},\label{eq:model}
\end{equation}
where $v_{F}$ is the Fermi velocity associated with the topological
surface state, $m^{*}$ is the effective mass of electrons(holes)
in the semiconductor, and $\Delta$ is the CBM(VBM) energy relative
to Dirac point, as shown in Fig. \ref{fig:Fig1}(b). For the most
general type of interaction, the $2\times2$ interaction matrix $V_{{\rm int}}$
can be expanded in terms of Pauli matrices, 
\begin{equation}
V_{{\rm int}}=\alpha_{0}\identity+\sum_{i=1}^{3}\alpha_{i}\sigma_{i},\label{eq:v_int}
\end{equation}
where each $\alpha$ is an independent complex parameter. In the absence
of interaction, the solutions to equation \eqref{eq:model} at the
$\Gamma$-point leads to two sets of doubly degenerate solutions;
those associated with the Dirac-point of the topological interface
state at $E=0$ and those of the doubly degenerate semiconductor state
at $E=\Delta$. This degeneracy at the $\Gamma$-point is the spin
degeneracy which is guaranteed by Kramers theorem, whereby operating
on the eigenfunction with the time reversal operator $\mathbf{T}=i\mathbf{K}\identity\otimes\sigma_{2}$
(where $\mathbf{K}$ represents complex conjugation) yields the eigenfunction
of the energy degenerate Kramers pair. It is the TRS which is said
to protect the topological state, as only perturbations which break
TRS (such as the presence of magnetic impurities) can break the degeneracy.

Herein, however, we focus our attention on perturbations which preserve
the TRS, as these are most relevant at the ``standard'' TI/NI interface
in the absence of localized magnetic moments or long range ordering.
This is accomplished by requiring that $V_{{\rm int}}$ commutes with
the time reversal operator, $\left[\left(\alpha_{0}\identity+\sum_{i=1}^{3}\alpha_{i}\sigma_{i}\right),\mathbf{T}\right]=0$
at the time reversely invariant momenta (TRIM) points in the Brillouin
zone (BZ). This leads to the following constraints for the $\alpha$-parameters:
\begin{equation}
\alpha_{0}=\alpha_{0}^{*},\ \ \ \alpha_{i}=-\alpha_{i}^{*},\ i=1,2,3,
\end{equation}
whereby $\alpha_{0}$ must be wholly real and $\alpha_{1}$, $\alpha_{2}$
and $\alpha_{3}$ are purely imaginary. Solving for the energy eigenvalues
of equation \eqref{eq:model} we find that indeed interactions of
this form preserve the double degeneracy at the $\Gamma$-point, with
eigenvalues at 
\begin{equation}
E_{\Gamma}={\frac{1}{2}}(\Delta\pm\sqrt{\Delta^{2}+4\alpha_{M}^{2})},\label{eq:E_gamma}
\end{equation}
where we have defined the total magnitude of the interaction, $\alpha_{M}=\sqrt{|\alpha_{0}|^{2}+{|\alpha_{1}|}^{2}+{|\alpha_{2}|}^{2}+{|\alpha_{3}|}^{2}}$.
Upon first glance, not much seems to have changed, with the interaction
increasing the energy between the Dirac point and the band edge from
$\Delta$ to $\sqrt{\Delta^{2}+4\alpha_{M}^{2}}$. However, the interaction
also leads to an avoided crossing off $\Gamma$, as shown schematically
in Fig. \ref{fig:Fig1}(d). While this interaction \textit{cannot}
in principle open a gap, as it is an avoided crossing between the
doubly degenerate semiconductor band and the non-degenerate TI state,
it necessarily leads to a region where the interaction causes the
interconversion of the character of the semiconductor bands and the
TI state. Note that the lower branches of the Dirac cone cannot directly
join with the upper branch near the Dirac point as they cannot cross
the semiconductor band and hence the lower branch of the TI state
becomes degenerate with the bulk derived state at $\Gamma_{a}$. As
the degenerate states at $\Gamma$ become close in energy, neither
of these degenerate states can be clearly distinguished as the Dirac-point
as both exhibit the evolution of topological states into bulk semiconductor
derived states.

In order to investigate the nature of these mixed states, particularly
how the helical nature of the topological interface state is altered
under the interaction, we investigate the spin-textures of the four
bands of the model. In the current basis, the spin-textures of the
Bloch eigenstates of the effective Hamiltonian are calculated by the
expectation values of the $4\times4$ matrices: 
\begin{equation}
\boldsymbol{\Sigma}_{i}=\frac{\hslash}{2}\identity\otimes\sigma_{i},\ \ i=1,2,3.
\end{equation}
While as suggested by equation \ref{eq:E_gamma}, the calculated band
structure depends only on the total magnitude of the interaction,
$\alpha_{M}$, the nature of the resulting spin-texture is found to
be dependent on the details of the interaction and the relative magnitudes
of the contributing $\alpha_{i}$'s. Spin textures for the four bands
for different choices for the non-vanishing interaction parameter
$\alpha_{i}$ are shown in Fig. 2.

\begin{figure}
\centering \includegraphics[width=0.45\textwidth]{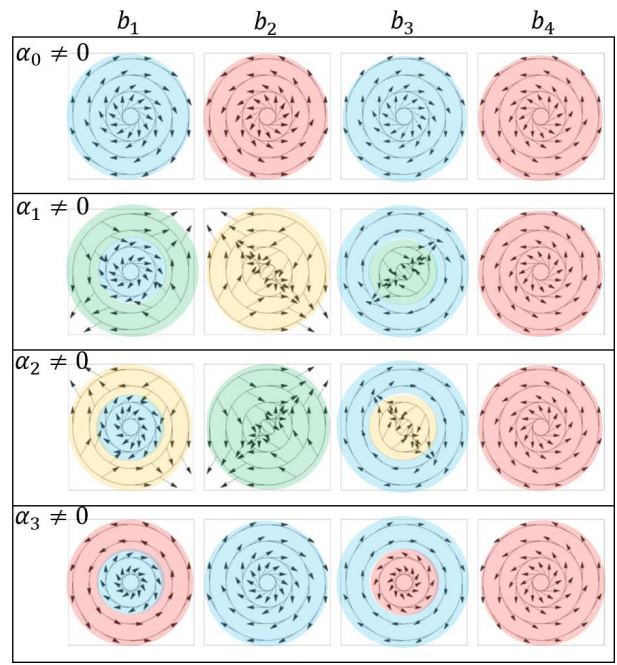} \caption{\textbf{Spin texture for the four bands of the model with different
non-vanishing interaction parameter $\alpha_{i}$ for an alignment
in which the semiconductor band lies below the Dirac-point at $\Gamma$}.
Each figure shows the spin orientation for a particular band for various
concentric circles in the $k_{x}$--$k_{y}$ plane, with the center
being the $\Gamma$-point. They are arranged in columns of differing
band number, labeled ($b_{1-4}$) in increasing energy, and in rows
for the different types of interaction. The blue and red shaded regions
indicate left and right handed spin texture, respectively. The yellow
and green regions indicated left and right handed anti-helical spin
texture, where the spin has a negative winding number.}

\label{fig:Fig2} 
\end{figure}

There are several novel features for these spin textures. Firstly,
for all cases, all four bands in the model become spin-textured as
a result of the interaction. In addition to the helical spin states
associated with the TI interface state, interactions containing $\sigma_{x}$
or $\sigma_{y}$ terms can induce an anti-helical spin texturing in
the bulk derived band. For these anti-helical states, see for instance
the yellow shaded $b_{2}$ under $\alpha_{1}$, as you transverse
the BZ in a counterclockwise direction the spin rotates in a clockwise
manner, leading to a winding number of -1. Also interesting is the
qualitative change in spin texture for individual bands, for instance
$b_{1}$ under $\alpha_{3}$ which changes from right-handed helical
to left-handed helical at some critical $|k|$. Although strange at
first sight, this has a straightforward explanation -- it corresponds
to a change of character of the bands in the vicinity of an avoided
crossing. Note that if the bulk derived state has a large curvature,
it is possible for it to intersect with the Dirac cone twice instead
of only once, leading to three distinct regions of spin texture. Although
this second critical $|k|$ is not guaranteed to exist, we mention
it here for the sake of completeness.

Focusing on the behavior close to $\Gamma$, we see that the highest
energy band, $b_{4}$, keeps its left-handed helical nature under
all interactions. This is expected as the semiconductor band intersects
with the Dirac cone below the Dirac point, hence the spin texture
of the top half of the Dirac cone ($b_{4}$) remains unchanged. Note
however, that $b_{3}$ which one might expect to be associated with
the bottom half of the Dirac cone (given that it is degenerate with
$b_{4}$ at $\Gamma$) does ${\it {not}}$ possess the right-handed
helical spin texture associated with the bottom half of the Dirac
cone, but instead can be either left- or right- helical or anti-helical.
Instead, it is found that only $b_{1}$ has the required right-handed
helicity.

spin texture associated with the bottom half of the Dirac cone, irrespective
of the interaction. This can be understood by referring back to schematic
Fig. 1. Here the Dirac cone is split, with the bottom and top half
of the Dirac cone no longer being degenerate at $\Gamma$, but instead
they become degenerate with the bulk derived semiconductor state,
still satisfying the required degeneracy due to TRS. In order to find
a system which may exhibit the exotic physical properties of the model,
we turn to the technologically relevant Bi$_{2}$Se$_{3}$/GaAs interface.

%\section*{First-principles Calculations}

\begin{figure}
\centering \includegraphics[width=0.45\textwidth]{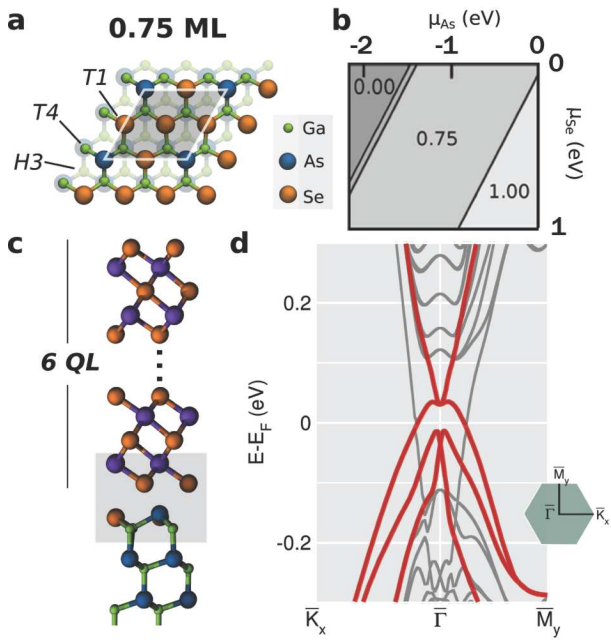} \caption{\textbf{First-principles calculations of Bi$_{2}$Se$_{3}$/GaAs interface.}
\textbf{(a)} Top-view of ball-and-stick model for GaAs$(111)$ surface
with 0.75 ML of Se-treatment. The unit cell is shaded in gray. The
stacking sites of the first atomic layer of Bi$_{2}$Se$_{3}$ are
labeled by $H3$, $T4$ and $T1$. \textbf{(b)} Chemical stability
map for GaAs$(111)$ surface. Are shown the lowest energy Se-treatment
in function of chemical potentials $\mu_{As}$ and $\mu_{Se}$. \textbf{(c)}
Side-view ball-and-stick model for Bi$_{2}$Se$_{3}$/GaAs interface
with 0.75 ML of Se-treatment and $H3$ stacking. The shaded area shows
the interface region. \textbf{(d)} Electronic band structure for Bi$_{2}$Se$_{3}$/GaAs
with 0.75 ML of Se-treatment. Four bands near the Fermi level are
highlighted in red for comparison with the effective Hamiltonian.}

\label{fig:Fig3} 
\end{figure}

In order to construct a realistic atomistic model of the Bi$_{2}$Se$_{3}$/GaAs
interface, we use First-principles density functional theory calculations
(see Methods) to first investigate the chemical stability and electronic
properties of the isolated Se-treated As-terminated GaAs$(111)$ surface.
In the $2\times2$ unit cell, we considered Se replacing between 0
to 4 surface As, yielding $0$ to $1.00$ monolayer (ML) of Se coverage:
$0$, $0.25$ ML, $0.50$ ML, $0.75$ ML, and $1.00$ ML. Ball-and-stick
models of geometries of GaAs$(111)$ surfaces with a Se-treatment
of $0.75$ ML is shown in Fig. \ref{fig:Fig3}(a). The chemical stability
map for the different Se coverages as a function of the chemical potentials
$\mu_{Se}$ and $\mu_{As}$ are shown in Fig. \ref{fig:Fig3} (b).
We focus our attention on the 0.75 ML coverage case as this is the
most energetically stable surface for a large range of chemical potentials
and leads to a fully passivated GaAs(111) surface which maintains
its bulk band gap.

\begin{figure*}[!ht]
\centering \includegraphics[width=0.98\textwidth]{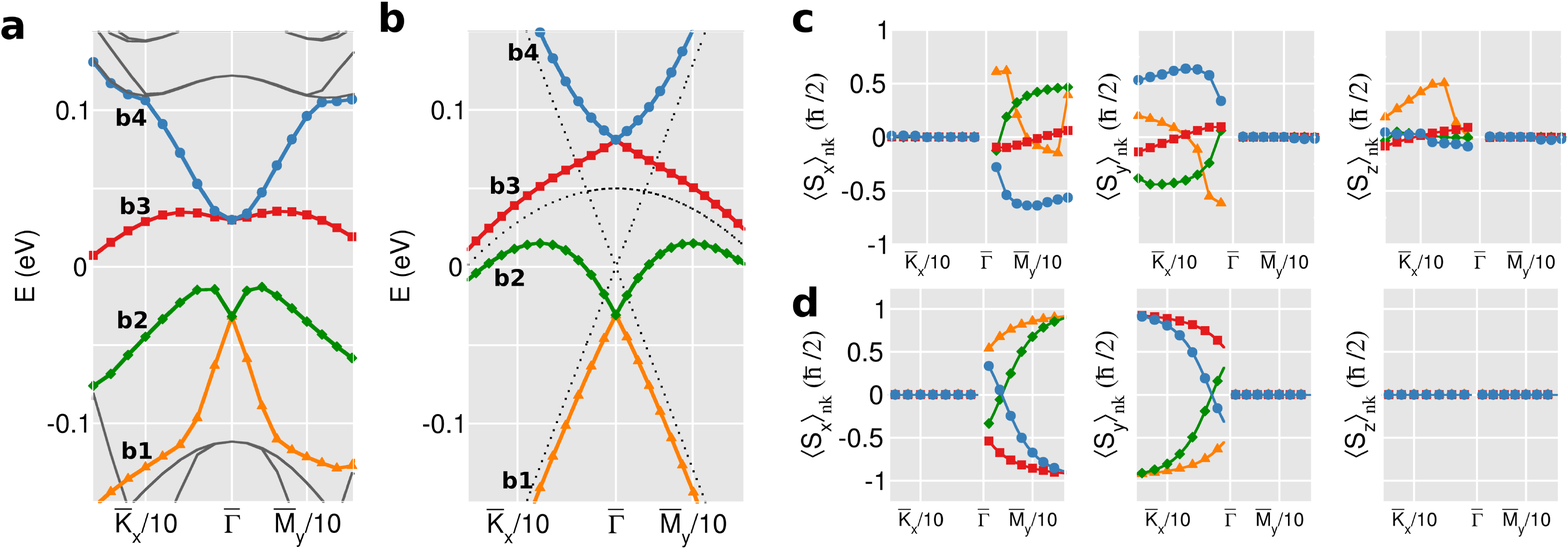} \caption{\textbf{First-principles calculations and Hamiltonian model for Bi$_{2}$Se$_{3}$/GaAs
interface}. \textbf{(a)} First-principles band structure of Bi$_{2}$Se$_{3}$/GaAs
for $0.75$ ML. \textbf{(b)} Effective model band structure for $\alpha_{3}=i0.05$
eV, $\alpha_{0}=\alpha_{1}=\alpha_{2}=0$, $v_{F}=5\cdot10^{5}$ m/s,
$m_{*}=-0.456m_{0}$ and $\Delta=0.05$. \textbf{(c)} First-principles
spin textures near $\overline{\Gamma}$-point for $b1$ (orange),
$b2$ (green), $b3$ (red) and $b4$ (blue). \textbf{(d)} Effective
Hamiltonian model spin textures near $\overline{\Gamma}$-point for
$b1$ (orange), $b2$ (green), $b3$ (red) and $b4$ (blue).}

\label{fig:Fig4} 
\end{figure*}

The interface is constructed by stacking $6$ QL of Bi$_{2}$Se$_{3}$
on the fully passivated 0.75ML Se substituted GaAs (111) surface,
as shown in \ref{fig:Fig3}(c), with the Se atomic layer of Bi$_{2}$Se$_{3}$
directly above the $H3$ substrate site indicated in Fig. \ref{fig:Fig3}
(a).The calculated band structure of the Bi$_{2}$Se$_{3}$/GaAs interface
is shown in Fig. \ref{fig:Fig3}(d). In order to identify which states
are localized at the interface and originate from the non-trivial
band topology, we perform a local real space projection of each of
the eigenfunctions onto the interfacial region shaded in Fig. \ref{fig:Fig3}
(c). The four bands which are most strongly localized to the interfacial
region are shown highlighted in red in the bandstructure. An enlargement
of the band structure near the $\overline{\Gamma}$-point and the
accompanying spin textures of these four bands, labeled $b1-b4$,
are shown in Fig. \ref{fig:Fig4}(a) and (c), respectively. These
four bands are directly comparable with the bands of the model, with
$b2$ and $b3$ originating from the VBM of GaAs and becoming degenerate
away from $\overline{\Gamma}$. Furthermore, the spin texture indicates
appreciable spin-locking in the $b1$, $b2$, and $b4$ bands.

To determine how representative the model presented in equation (1)
is of the physics at the interface, we independently determine the
material parameters for the model from first-principles calculation.
We find a Fermi velocity of $v_{F}=5\cdot10^{5}$ m/s for the Dirac
cone of the pristine Bi$_{2}$Se$_{3}$, an effective mass of $m^{*}=-0.456m_{0}$
for the heavy-holes of the GaAs surface, and the relative VBM energy
for the Bi$_{2}$Se$_{3}$/GaAs interface from workfunction calculations
of $\Delta=0.05$ eV. Using these parameters, we fit the single parameter
associated with the interaction $\alpha_{M}=0.05$ eV to produce the
model bandstructure shown in Fig. \ref{fig:Fig4} (b) to compare with
the ab-initio bandstructure in Fig. \ref{fig:Fig4} (a). By analyzing
the spin textures of the first-principles calculations, we find that
qualitative agreement can be reached when interaction is of the form
$V_{{\rm int}}\propto\sigma_{z}$, and hence we set $\alpha_{0}=\alpha_{1}=\alpha_{2}=0$
and $\alpha_{3}=i0.05$ eV.

The spin texture of the model Hamiltonian is shown in Fig. \ref{fig:Fig4}
(d). The agreement of the features of band $b2$ and $b4$ compare
quite well between the calculation and model, being primarily right
and left-handed helical states, respectively, with the spin-momentum-locking
of both greatly diminished in the immediate vicinity of $\Gamma$.
Bands $b1$ and $b3$, however, appear less well described. While
the nearly quenched spin-momentum-locking of $b3$ from ab-initio
calculation suggests that this band is less localized at the interface
(and consequentially has less coupling), the origin of the deviation
of $b1$ from the model is clear. As we move away from $\Gamma$ the
resulting decrease in the magnitude of spin is concurrent with the
interaction of $b1$ with lower lying bands which are not considered
in the model. The spin-locking once again becomes more pronounced
as the energy separation with lower lying bands increases. This suggests
that including more bands in the model will lead to a more quantitatively
accurate description. Nonetheless, we note that the essential feature
of the model is also found in the DFT calculation. That being that
the top-half of the Dirac cone, associated with left-handed helicity,
and the bottom half of the Dirac-cone, associated with right-handed
helicity, are separated by an energy gap at $\Gamma$.

The model presented in eqn. (1) and the finding of its applicability
to realistic interfaces suggest that not only is the particle-hole
symmetry of the Dirac cone fundamentally broken at the interface,
but that a number of new emerging phenomenon with potential applications
exist. Firstly, while optical excitations from linearly polarized
light are spin forbidden for an ordinary Dirac cone, the spin interaction
present at the interfacial Dirac cone relaxes this selection rule,
leading to a number of allowed transitions which generally become
$|k|$-dependent. This opens up the possibility of directly probing
the topological interface states through optical means, such as absorption
or photoluminescence. Furthermore, the well delineated regions of
different spin texture shown in Fig. 2 indicate the presence of a
${\it spin}$-bottleneck. As electron relaxation into regions with
a different spin texture would require a magnetic interaction, electron
relaxation is now confined to individual shells of different spin
textures with a prolonged excitation lifetime. In this way, it is
similar to the phonon bottleneck which limits the relaxation rates
of hot carriers in graphene \cite{fwang2014nanolett}, potentially
leading to photosensitive conductivity and the prospect of the formation
of an excitonic condensate at the interface. %\section*{Conclusion}

In summary, we present a general and simple model to describe the
interaction of the interfacial Dirac-cone with non-topological states
at the interface. The main features of the model are demonstrated
in our parameter-free density-functional theory calculations of the
Bi$_{2}$Se$_{3}$(0001)/GaAs$(111)$ interface. This interaction
is found to lead to a number of emerging phenomenon which highlight
the differences between the Dirac cone at the surface of a TI and
that found at the interface. These includes the splitting of the Dirac
cone (without breaking TRS), the acquisition of spin-texture of bulk
derived bands, alteration of optical selection rules, the fundamental
breaking of particle-hole symmetry, and the presence of a spin-bottleneck
with associated long excitation lifetimes. While we have focused on
the interaction of the emerging Dirac cone with bulk derived states
at the TI/NI interface, the model is more general and could be trivially
extended to investigate the interactions of the surface Dirac-cone
with other non-topological states. These findings both suggest new
possible phenomenon at the interface and may help to guide developments
in TI/NI-based spintronics devices, especially due to the helical
topological states role in spin current generation\cite{pesin2012spintronics,li2014electrical}
and spin relaxation\cite{zutic2004spintronics}.

\section*{Methods}

The atomistic first principles calculations were performed within
the Density Functional Theory (DFT) framework\cite{PhysRev.136.B864,PhysRev.140.A1133}
as implemented in the \textit{Vienna Ab Initio Simulation Package}
(\textsc{Vasp})\cite{PhysRevB.54.11169,PhysRevB.59.1758}. Energy
cutoff of 300 eV and $k$-points grid of $3\times3\times1$ in Monkhorst--Pack
algorithm\cite{PhysRevB.13.5188} were taken as default on our calculations.
The external potential was calculated using Projector Augmented-Wave
(PAW) method\cite{PhysRevB.50.17953,PhysRevB.59.1758}. The exchange-correlation
functional was in Local Density Approximation (LDA)\cite{PhysRevLett.45.566}
parameterized by Perdew--Zunger\cite{PhysRevB.23.5048} due to agreement
with experimental bond lengths among Bi$_{2}$Se$_{3}$ quintuple-layers\cite{seixas2013topological}.
Electronic band structures and spin textures required non-collinear
spin polarizations for spin-orbit interactions. The chemical stability
of Se on the isolated As terminated GaAs(111) surface was determined
by slab calculations in a 2x2 unit cell with 18 atomic layers. The
Ga-terminated surface was passivated with H with a fractional charge
(q=1.25 e). In order to minimize the effect of strain, a lattice constant
of 4.07 $\AA$ was used in the interface calculation. The stacking
with the Se atomic layer of Bi$_{2}$Se$_{3}$ directly above the
$H3$ substrate was chosen as it was found to be 20 and 490 meV/supercell
lower in energy than the $T4$ and $T1$ configurations, respectively.

\section*{Authors contributions}

L.S. performed the first principles calculations. L.S. and D.W performed
effective Hamiltonian model calculations and wrote the manuscript
with contributions from all authors. A.F. supervised the effort in
Brazil and S.B.Z. conceived the project. All authors equally discussed
the results and contributed to the understanding.

\section*{Acknowledgments}

L.S. and A.F. acknowledges the financial support by the Brazilian
agencies CNPq/INCT and FAPESP. L.S. also acknowledges partial financial
support provided by Rensselaer Polytechnic Institute. D.W. acknowledges
support from the Defense Award Research Project Agency, Award No.
N66001-12-1-4304, and S.B.Z. acknowledges support from the US Department
of Energy under Grant No. DE-SC0002623. Supercomputer time was provided
by NERSC under Grant No. DE-AC02-05CH11231 and the Center for Computational
Innovations at Rensselaer Polytechnic Institute.

%\bibliography{ti}
% \bibliographystyle{naturemag} %\begin{thebibliography}{50}
%\end{thebibliography}

\expandafter\ifx\csname url\endcsname\relax \global\long\def\url#1{\texttt{#1}}
\fi \expandafter\ifx\csname urlprefix\endcsname\relax\global\long\def\urlprefix{URL }
\fi \providecommand{\bibinfo}[2]{#2} \providecommand{\eprint}[2][]{\url{#2}}

\end{document}